\documentclass[prb,twocolumn]{revtex4}
    \usepackage{amssymb}
    \usepackage{graphicx}
    \usepackage{amsfonts}              
    \usepackage{amsmath}
    \usepackage{bm}
    \usepackage{comment}

\begin{document}
    \title{Topological spin-Hall current in waveguided zinc-blende semiconductors with Dresselhaus spin-orbit coupling}   
    \author{T. Fujita}       
    \affiliation{Information Storage Materials Laboratory, Electrical and Computer Engineering Department, National University of Singapore, 4 Engineering Drive 3, Singapore 117576}
    \affiliation{Data Storage Institute, DSI Building, 5 Engineering Drive 1, (off Kent Ridge Crescent, National University of Singapore) Singapore 117608}
   
   \author{M. B. A. Jalil}
   \affiliation{Information Storage Materials Laboratory, Electrical and Computer Engineering Department, National University of Singapore, 4 Engineering Drive 3, Singapore 117576}
    \author{S. G. Tan}
   \affiliation{Data Storage Institute, DSI Building, 5 Engineering Drive 1, (off Kent Ridge Crescent, National University of Singapore) Singapore 117608}
    
\begin{abstract}
We describe an intrinsic spin-Hall effect in $n$-type bulk zinc-blende semiconductors with topological origin. When electron transport is confined to a waveguide structure, and the applied electric field is such that the spins of electrons remain as eigenstates of the Dresselhaus spin-orbit field with negligible subband mixing, a gauge structure appears in the momentum space of the system. In particular, the momentum space exhibits a non-trivial Berry curvature which affects the transverse motion of electrons anisotropically in spin, thereby producing a finite spin-Hall effect. The effect should be detectable using standard techniques in the literature such as Kerr rotation, and be readily distinguishable from other mechanisms of the spin-Hall effect.
\end{abstract}


\maketitle

\section{Introduction}
The spin-Hall effects (SHE) are a set of phenomena in which a transverse spin current is generated in response to an applied electric field \cite{engel}. Originally predicted by Dyakonov and Perel' \cite{dyakanov-perel} more than three decades ago, and later revisited by Hirsch \cite{hirsch}, the SHE has developed into a topic of keen interest and importance in the condensed matter field of semiconductor-based spintronics, for it allows one to generate and manipulate spin currents in paramagnetic semiconductors without the application of external magnetic fields or the use of ferromagnetic components. A key ingredient in the theory of SHEs is the spin-orbit coupling (SOC) effect; a phenomenon that is formally described by Dirac's equation when decomposed in the non-relativistic limit, in which a static electric field $\vec{E}=\nabla V$ is Lorentz transformed into an effective magnetic
field $\vec{k}\times\nabla V$ in the rest frame of moving electrons ($\vec{k} $ is the electron momentum). In Refs.\ \cite{dyakanov-perel,hirsch} it is  predicted that the SOC of carrier momentums with the localized potential centres of impurity atoms results in spin-dependent scattering of the carriers. The essential result is that spin-up and spin-down electrons are scattered in opposite transverse directions, resulting in a so-called extrinsic (impurity-dependent) SHE \cite{mott, berger}. More recently there has been widespread interest in the study of intrinsic SHE mechanisms, seeded by two seminal papers; Ref.\ \cite{murakami} which describes the transverse spin transport of holes in $p$-type bulk semiconductors, and Ref.\ \cite{sinova} in $n$-type two dimensional heterostructures with Rashba SOC. In contrast to the extrinsic type, the intrinsic SHE does not depend on the SOC between carriers and impurities, but rather on the `built-in' SOC that is present in the band structure of the system.
In Ref.\ \cite{murakami}, the strong spin-orbit interaction in the valence band of bulk semiconductors was shown to give rise to a nontrivial momentum space topology under an applied electric field, resulting in the flow of a topological spin-Hall current. On the other hand, in Ref.\ \cite{sinova}, the Rashba SOC produces a SHE that is a result of the spin precession about the internal Rashba field in the presence of an electric field. 
We focus on the topological SHE of Ref.\ \cite{murakami}. The SHE described there is induced by the presence of a magnetic monopole field in $\vec{k}$-space, that arises from the coupling of the orbital angular momentum $\vec{L}$ and spin angular momentum $\vec{S}$ of holes in the valence band under the influence of an externally applied electric field. The presence of the monopole results in non-commuting coordinates \cite{murakami,bliokh}, which leads to a Lorentz-type force in $\vec{k}$-space and a separation of spins leading to the SHE.\\
The appearance of the monopole structure is not unique to $p$-doped, bulk semiconductors as studied in Ref.\ \cite{murakami}. It also appears in other condensed matter systems, such as in the anomalous Hall effect (AHE) in ferromagnets \cite{onoda-nagaosa,fang}, the AHE in frustrated ferromagnets---e.g.\ pyrochlore and Kagom\'{e} lattices---with chiral spin textures \cite{taguchi,ye,chun}, and in the so-called topological Hall effect in specially patterned magnetic nanostructures \cite{bruno,bruno2}. The monopole in general appears through the nontrivial curvature of gauge fields that are associated with an adiabatically evolving quantum system \cite{berry1983,simon}. For example, assuming a slowly varying magnetic field configuration over a parameter space $\Pi$, one can impose the condition of adiabatic spin relaxation (i.e.\ that the spins remain as eigenstates of the field), which in accordance with Ref.\ \cite{berry1983} gives rise to a Berry curvature (the monopole) in $\Pi$-space.
When dealing with the momentum space, i.e.\ $\Pi = \vec{k}$, the curvature can be regarded as an effective, momentum dependent field which can influence the motion of carriers (analogous to an ordinary field in real space), leading to modified carrier dynamics. It should be noted, however, that the appearance of a monopole curvature does not automatically result in a spin-Hall effect; this depends on other details of the system.\\
In this paper, we describe an intrinsic spin-Hall effect of conduction electrons in $n$-type bulk semiconductors with $k^3$-Dresselhaus SOC that are confined by a waveguide to propagate primarily in a unilateral direction. By applying an external electric field along the direction of the waveguide, we show how a monopole structure appears in the momentum space and leads to a finite SHE. We consider the system in the weak applied field limit such that the internal Dresselhaus field is smoothly varying, and the spins remain adiabatically aligned along its direction. We discuss the origin of our SHE in detail and finally propose experimental setups that may be used to detect the effect, as well as to distinguish the effect from other mechanisms.
\section{Theory}
\subsection{Appearance of gauge structure}
\begin{figure}[ht!] 
    \begin{center} 
    \resizebox{0.9\columnwidth}{!}{ 
    \includegraphics{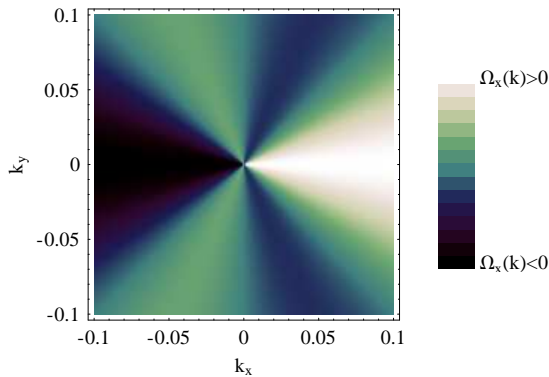}} 
    \end{center}
    \caption
    {(color online). Distribution of the $\Omega_x(\vec{k},s=+1)$ field component of the Berry curvature in $\vec{k}$-space described by Eq.\ \eqref{mono}. For simplicity, we use normalized values for the momentum, $k_z =1$ and $|k_x,k_y| \le 0.1$. The lighter (darker) regions correspond to increasingly positive (negative) values of $\Omega_x(\vec{k})$. The unusual behavior near $k_x=k_y=0$ is a direct consequence of the Dirac monopole at the origin of $\vec{B}_D(\vec{k})$-space. Both $\Omega_x(\vec{k})$ and $\Omega_y(\vec{k})$ play important roles in driving the spin-Hall effect presented in this paper. The figure for $\Omega_y(\vec{k})$ is identical to the one above, but with $k_x$ and $k_y$ axes interchanged.}
    \label{field.fig}
    \end{figure}

We consider the Dresselhaus spin-orbit coupling in the conduction band of bulk zinc-blende semiconductors. The conduction electrons in this system are described by the Hamiltonian 
\begin{eqnarray}
	\mathcal{H}&=&\hbar^2 k^2/2m + \mathcal{H}_D + V(\vec{r}), \text{ where}\label{tothamil} \\
	\mathcal{H}_D & =&
	\eta \left(\sigma_x k_x(k_y^2-k_z^2)+\text{c.p.}\right)\nonumber \\ 
	& \equiv & \eta \vec{\sigma}\cdot\vec{B}_D \label{hamiltonian},
\end{eqnarray}
and $\eta$ is the Dresselhaus SOC strength (units eVm$^3$), $\vec{\sigma}$ is the vector of Pauli matrices, $k_i$ are the electron momenta along $i=[\vec{e}_i]$ of the crystal, c.p.\ denotes the cyclic permutation in $i$ of the spin-orbit term, and $\vec{B}_D=\vec{B}_D(\vec{k})$ is a momentum-dependent internal magnetic field \cite{dressel}. The last term in \eqref{tothamil}, $V(\vec{r})=e \vec{E}\cdot\vec{r}$, is the potential energy of electrons due to external electric field $\vec{E}$. The corresponding eigenvalues of \eqref{tothamil} are $E_s =E_0 +s \eta |\vec{B}_D(\vec{k})|+V(\vec{r})$ where $E_0$ ($V$) is the kinetic (potential) energy and $s=\pm$ indexes the two spin-split subbands of the Dresselhaus Hamiltonian.\\
We begin our analysis by applying a local, unitary transformation $U=U(\vec{k})$ to the system, such that the reference spin axis points along the direction of $\vec{B}_D(\vec{k})$. Under this transformation the Hamiltonian becomes diagonalized, $U\mathcal{H}U^\dagger=\hbar^2 k^2/2m + s\eta \sigma_z |\vec{B}_D|+U V(\vec{r}) U^\dagger$, in the spinor space. In the momentum representation, the potential energy term transforms as $U V(\vec{r}=i\nabla_k) U^\dagger=e\vec{E}\cdot(i\nabla_k - i U\nabla_k U^\dagger)$. Here, the position operator $\vec{r}$ transforms into covariant form: $\vec{r} \rightarrow \vec{R}=\vec{r}+\vec{A}(\vec{k})$, where $\vec{A}(\vec{k})=-iU\nabla_k U^\dagger$ is a gauge field in $\vec{k}$-space \cite{murakami}. There is a clear analogy (and duality) here with standard electromagnetism: in the presence of an external magnetic field $\vec{B}$, the momentum operator $\vec{p}$ transforms as $\vec{p}\rightarrow \vec{\Pi}=\vec{p}+\vec{A}(\vec{r})$, where the last term is the magnetic vector potential (whose curvature equals $\vec{B}$). The computation of the gauge field $\vec{A}(\vec{k})$ in our system was carried out firstly by assuming that the travelling wavevector, $k_z$, (we consider an applied electric field in the $\hat{z}$-direction) has magnitude greater than the transverse wavevector components, $k_\parallel=(k_x,k_y)$. In other words, we assume electron conduction primarily along the $\hat{z}$-direction, with a minimal angular spectrum of electrons in the transverse $\hat{x}\hat{y}$-plane. This can be achieved with the help of a confinement potential $V(x,y)$ which confines electrons to a waveguide along $\hat{z}$; see, for example, Ref.\ \cite{datta}. Under the assumption $k_z \gg k_x,k_y$, the effective internal field  can be approximated by the simplified field $\vec{B}_D^{\text{ }'}(\vec{k})=(-k_x k_z^2,k_y k_z^2, k_z(k_x^2-k_y^2))$ \footnote{One may recognize that $\vec{B}_D^{\text{ }'}(\vec{k})$ is also the effective internal field within a tunnel barrier with  $k$-cubic Dresselhaus SOC; see, for example, Ref.\ \cite{perel}. The condition $k_z \gg k_x, k_y$ is fulfilled automatically in this scheme, i.e.\ when the kinetic energy of electrons lies much lower than the potential barrier height \cite{perel}, and need not be imposed explicitly as in our case which requires the use of waveguides \cite{datta}. Numerically, we verified that the approximation is very good when the condition $k_z \gtrsim 3 \sqrt{k_x^2+k_y^2}$ is met.} (hereafter, for
brevity, $\vec{B}_D^{\text{ }'}(\vec{k})$ shall be denoted by $\vec{B}_D(\vec{k})$). Then, $\vec{A}(\vec{k})$ can be found readily using the explicit expression for the transformation, $U(\vec{k})=\exp{(-i \frac{\theta}{2} \vec{\sigma}\cdot\vec{n})}$, where $\vec{\sigma}\cdot\vec{n}=\sigma_x \sin{\phi}-\sigma_y \cos{\phi}$ and $\theta,\phi$ are spherical angles satisfying $\cos{\theta} = B_{Dz}/|\vec{B}_D|$ and $\tan{\phi} = B_{Dy}/B_{Dx}=-k_y/k_x$ respectively. Unlike the case for electromagnetism, however, our computed gauge field is pure and has no associated curvature. Nevertheless, upon imposing the adiabaticity condition for the spins, we can induce a finite curvature. More specifically, we neglect subband mixing due to the electric field, i.e.\ we suppose that as electrons drift through the crystal under $\vec{E}$, their spins remain as eigenstates of the effective $\vec{k}$-dependent magnetic field $\vec{B}_D(\vec{k})$, and that transitions between the two eigenstates (up-spin and down-spin) are negligible.  Generally speaking, this condition can be realized in quantum systems when the Hamiltonian is varied sufficiently smoothly (adiabtically) over time via one of its parameters \cite{simon}. The adiabatic condition in our system corresponds to the low applied $\vec{E}$-field limit, such that changes in time of the electron momentum---and hence, the effective magnetic field---are sufficiently small so that adiabatic spin relaxation may be realized. We estimate the value for the required $\vec{E}$-field in a latter part of this paper.\\
Assuming adiabaticity of the electron spins, we can throw away the off-diagonal (transition) terms of $\vec{A}(\vec{k})$. The resulting spin-state resolved U(1)-gauge fields have the form $\vec{A}(\vec{k},s)=is/2(1-\cos{\theta}) \nabla_k \phi$, and have a finite curvature that is the Berry curvature in momentum space,
\begin{equation}
F_{ij}(\vec{k},s)\equiv\Omega_k(\vec{k},s)=\partial_i A_j - \partial_j A_i =s\epsilon_{ijk} \frac{k_z^4 (k_x^2-k_y^2)}{2 |\vec{B}_D(\vec{k})|^3}k_k,
\label{mono}
\end{equation}
where $\epsilon_{ijk} $ is the fully asymmetric tensor in three dimensions.  As expected \cite{berry1983}, the curvature in Eq.\ \eqref{mono} exhibits singularities  at points where $|\vec{B}_D(\vec{k})|=0$, corresponding to the degeneracy points of the spin-dependent Hamiltonian in Eq.\ \eqref{hamiltonian}. Our curvature term above can also be derived in the spirit of Berry's original paper \cite{berry1983}, in which we first diagonalize the system Hamiltonian with respect to the magnetic field space $\vec{B}_D$. In doing so, we obtain Berry's curvature in $\vec{B}_D$-space that is of the form of the Dirac monopole, $F_{ij}(\vec{B}_{D},s)=s \epsilon_{ijk} B_{Dk}/|\vec{B_D}|^3$. The corresponding curvature in $\vec{k}$-space can then be found using the explicit dependence of the effective field on the electron momentum \cite{bliokh}.\\
In Fig.\ \ref{field.fig}, we illustrate one component of the Berry curvature, $\Omega_x(\vec{k},s=+1)$, for normalized values of momentum, $k_z=1$ and $-0.1\le k_x,k_y \le 0.1$. One can see that the curvature term appears smooth at all points except at the origin $k_\parallel = 0$ where the limit is undefined. This feature of our curvature can be attributed directly to the existence of the singularity of $\vec{F}(\vec{B}_{D})$ in the effective field space, and leads to the non-trivial electron dynamics which characterizes our SHE.
\subsection{Modified equations of motion}
\begin{figure}[ht!] 
    \begin{center} 
    \resizebox{0.9\columnwidth}{!}{ 
    \includegraphics{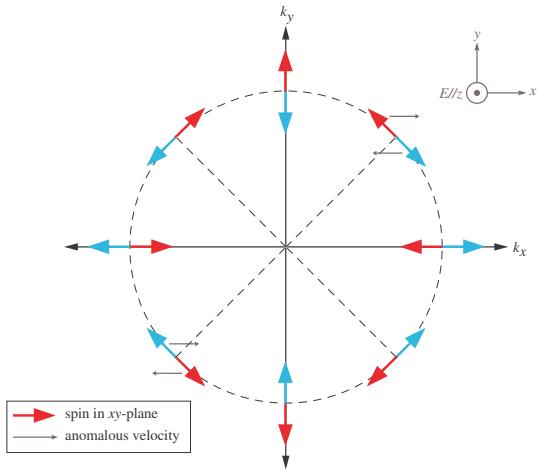}} 
    \end{center}
    \caption
    {(color online). Illustration of spin-Hall mechanism via anomalous velocity of electrons with momenta $(k_x,k_y)$ in a bulk Dressselhaus spin-orbit coupled system, under applied electric field in $\hat{z}$ direction. The spin orientations in the azimuthal plane are shown for the $+$ subband (red or dark gray arrows) and $-$ subband (cyan or light gray arrows). The anomalous velocity in the $x$-direction (horizontal, black arrows) due to the topological field in $k$-space for two values of momentum are shown. One can see that electrons with spin polarized along $+y$ ($-y$) experience an anomalous velocity in the $+x$ ($-x$)-direction. Although not shown, this applies for all electron modes over the Fermi circle, resulting in a finite spin current $j_x^y$ in the sample.}
    \label{directions.fig}
    \end{figure}

Albeit in reciprocal space, Eq.\ \eqref{mono} represents a field that is analogous to an ordinary magnetic field in real space in non-commutative quantum mechanics since in the presence of our gauge one can show such relations as $[r_i,r_j]=-i F_{ij}(\vec{k})\equiv-i\Omega_k (\vec{k})$. In contrast, for a classical magnetic field $\vec{B}$ with vector potential $\vec{A}$, i.e.\ $\text{curl }\vec{A}=\vec{B}$, the canonical momentums become non-commuting in a similar way, namely we have the relations $[p_i,p_j]=-ie B_k(\vec{r})$ from which the Lorentz force follows from Heisenberg's equation of motion, i.e.\ $\vec{F}=m/i\hbar [\vec{v},\mathcal{H}]$. The apparent underlying duality allows $\vec{\Omega} (\vec{k})$ to be interpreted as a magnetic field in momentum space, which gives rise to a $\vec{k}$-space `Lorentz-type' force. Just like a classical field, $\vec{\Omega} (\vec{k})$ affects the motion of iternerant electrons, which is characterized by the equations of motion derived in Ref.\ \cite{sundaram-niu}:
 \begin{eqnarray}
 	\hbar \dot{\vec{k}}&=&-e\vec{E} \label{eqm1}\\
 	\dot{\vec{r}}&=&\frac{1}{\hbar}\frac{d E_s}{dk}-\dot{\vec{k}}\times \vec{\Omega}(\vec{k})
 	\label{eqm2}
 \end{eqnarray}
 The last term in Eqn.\ \eqref{eqm2} is the Karplus-Luttinger anomalous velocity term \cite{karp-lutt} which has recently been used to describe intrinsic spin-Hall effects in doped semiconductors \cite{murakami} as well as to explain the anomalous Hall effect in ferromagnets \cite{fang,onoda-nagaosa}. Solving the above coupled equations of motion by integration we yield the real space trajectory of conduction electrons within our semiconductor system:\footnote{The first part of Eq.\ \eqref{eqm2}, the group velocity of the wavepacket, contains spin-dependent terms but these do not contribute to the spin-Hall current and have therefore been neglected in Eqs.\ \eqref{xyz} for simplicity. See under the heading \emph{Spin-Hall conductivity} for more details.}
 \begin{subequations}
\begin{eqnarray}
x(t)&=&x_0+\frac{\hbar k_{x0}}{m}t+s k_{y0} \gamma\label{x}\\
y(t)&=&y_0+\frac{\hbar k_{y0}}{m}t-s k_{x0} \gamma \label{y}\\
z(t)&=&z_0+\frac{\hbar k_{z0}}{m}t - \frac{\hbar e E_z}{2m}t^2 \label{z}
\end{eqnarray}
\label{xyz}
\end{subequations}
where the $0$-subscript denotes values at $t=0$, and
\begin{equation}
\gamma=-\frac{e E_z}{\hbar}\int_0^t \frac{k_z(\tau)^3 B_{Dz}(\vec{k}(\tau))}{2|\vec{B}_D(\vec{k}(\tau))|^3}d\tau,
\end{equation}
where $\vec{k}(\tau)=(k_{x0},k_{y0},k_z(\tau))$ and $k_z(\tau)=k_{z0}-e E_z \tau/\hbar$. Eq.\ \eqref{z} describes the drift motion of electrons under the applied electric field $\vec{E}=E_z \hat{z}$, whilst Eqs.\ \eqref{x} and \eqref{y} describe the motion of electrons in the plane perpendicular to the principal electron motion along $\hat{z}$. From Eqs.\ \eqref{x} and \eqref{y}, an electron experiences an anomalous velocity in the $\hat{x}\hat{y}$-plane that is perpendicular to its lateral momentum $(k_{x0},k_{y0})$, and whose exact direction (i.e.\ the sign) is governed by the subband $s$ the electron belongs to. The anomalous velocity term actually results in a finite SHE as we discuss below. Each subband of the Dresselhaus Hamiltonian comprises of an ensemble of degenerate modes whose spins are calculated via the expectation value of the Pauli spin operators. For illustration, let us focus on the $\hat{y}$-spin components.
 In the cubic Dresselhaus Hamiltonian that we use, one can show that for an eigenstate $|\vec{k},s\rangle$ of the system, $\langle \vec{k},s | \sigma_y | \vec{k},s \rangle = s k_y g(\vec{k})$ where $g(\vec{k})$ is a scalar-valued function with $g(\vec{k}) \ge 0$ over the entire $\vec{k}$-space. From Eq.\ \eqref{x}, we find that the anomalous velocity component along $\hat{x}$ has opposite signs for net positive $\langle \sigma_y \rangle$ and net negative $\langle \sigma_y \rangle$. Evidently, this gives rise to a finite spin current polarized along $\hat{y}$ and flowing in the $\hat{x}$-direction of the sample, i.e.\ $j_x^y \neq 0$, and explains the origin of our SHE. In Fig.\ \ref{directions.fig} we show the spin orientations for different values of the lateral momentum $(k_x,k_y)$ for the $+$ subband (red or dark gray arrows) and the $-$ subband (cyan or light gray arrows), and indicate the direction of the anomalous velocity experienced by electrons in the $x$-direction (horizontal black arrows) for two values of momentum. Although not shown, all spins pointing along the positive $\hat{y}$ direction experience an anomalous velocity along $+\hat{x}$, and vice-versa, resulting in a separation of spins polarized along $\hat{y}$ in the $\hat{x}$-direction of the sample. Due to symmetry in the transverse plane, there is also a spin current $j_y^x$ polarized along $\hat{x}$ flowing in the $\hat{y}$-direction of the sample, so we have a rotationally invariant spin current which can be characterized by the response equation \cite{murakami}
\begin{equation}
j_j^i=\sigma_s \epsilon_{ijk} E_k,
\label{shc.eq}
\end{equation}
where $\sigma_s$ is the spin-Hall conductivity.
\subsection{Spin-Hall conductivity}
\begin{figure}[ht!] 
    \begin{center} 
    \resizebox{\columnwidth}{!}{ 
    \includegraphics{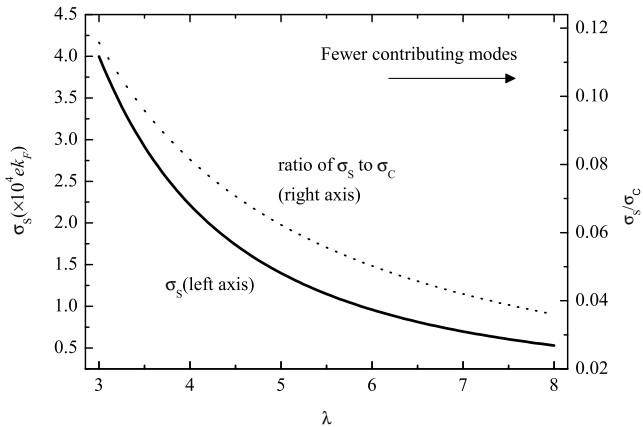}} 
    \end{center}
    \caption
    {(Left axis) Spin-Hall conductivity $\sigma_s$ as a function of $\lambda$. As fewer modes contribute to the spin-current, $\sigma_s$ diminishes. To compare with the charge conductivity we also plotted the ratio $\sigma_s/\sigma_c$ against $\lambda$ (right axis), and find that this ratio does not roll off as quickly as $\sigma_s$ itself. The spin-Hall current can therefore be fairly prominent against the longitudinal charge conduction for larger values of $\lambda$.}
    \label{spin_cond.fig}
    \end{figure}

To calculate the spin-Hall conductivity in Eq.\ \eqref{shc.eq} we use a semi-classical approach \cite{sinova,murakami}, and engage the conventional definition of the spin current operator in the $\hat{x}$-direction, $j_x ^y = \frac{1}{2}\langle\{s_y, v_x\}\rangle$, performing a summation over all states up to the Fermi level, $\vec{k}=\vec{k}_F$. We note that in the presence of spin-orbit coupling, the velocity operator in the $\hat{x}$ direction, $v_x$, contains spin-dependent terms from the Hamiltonian; namely we have $v_x=\partial E(\vec{k})/\partial k_x=\hbar k_x/m - k_z^2 \sigma_x + 2k_z k_x \sigma_z$ from Hamilton's equation, but these vanish in the anticommutator with $\sigma_y$ and therefore do not contribute to the spin-Hall current. Assuming that the spin splitting $\Delta = 2\eta |\vec{B}_D|$ from the Dresselhaus SOC is much smaller than the kinetic energy of electrons $E_0$ (valid for typical materials and doping densities: e.g.\ using data from Ref.\ \cite{kato} for $n$-GaAs, $E_0=5.7$meV $\gg 0.05$meV) we approximate the Fermi surface to be a 2-sphere in $\vec{k}$-space. For our waveguide channel, where the transport is primarily unilateral along $\hat{z}$, the region of interest $\mathcal{K}$ of the Fermi surface is the cap defined by $\mathcal{K}=\{\vec{k}: k_z \ge \lambda \sqrt{k_x^2+k_y^2} \text{ and } |\vec{k}|=k_F\}$, where $\lambda>1$ depends on details of the confinement potential $V(x,y)$. The value of the spin current is
\begin{eqnarray}
j_x^y & = & \sum _{s=\pm,\vec{k}\in \mathcal{K}} \dot{x}(\vec{k},s) \langle s_y(\vec{k},s) \rangle \\
& = & \sum_{s=\pm,\vec{k}\in\mathcal{K}} \frac{e E_z}{\hbar} \Omega_y(\vec{k},s)\langle \vec{k},s | \frac{\hbar}{2}\sigma_y | \vec{k},s \rangle \\
& = & \frac{e E_z}{2} \sum_{s=\pm} \int_\mathcal{K} \frac{d^3 k}{(2 \pi)^3} \Omega_y(\vec{k},s)\langle \vec{k},s | \sigma_y | \vec{k},s \rangle \label{response}
\end{eqnarray}
from which we obtain a spin-Hall conductivity of $\sigma_s \equiv j_x^y/E_z=4 \times 10^{-4} e k_F$ for $\lambda=3$. To generalize, we plot $\sigma_s$ (normalized to $e k_F$) as a function of the parameter $\lambda$ in Fig.\ \ref{spin_cond.fig}. We find that as $\lambda$ is increased, which corresponds to restricting the Fermi surface to smaller caps, the spin-Hall conductivity decays exponentially. It should be noted that for simplicity, we ignored any quantization effect from $V(x,y)$ in our calculations.
\subsection{Adiabaticity criterion}
As alluded to previously, our effect arises in waveguides of $n$-doped bulk zinc-blende materials in the adiabatic limit. In this limit the spins follow the quantization axis set in the direction of the effective magnetic field, allowing one to apply the Abelian approximation and to obtain a non-vanishing Berry curvature \eqref{mono}. Following \cite{bruno}, the adiabaticity condition is satisfied when the rate of change of the spin-quantization axis, $R_q$, is much smaller than the Larmor precession frequency. This guarantees that the spins have time to relax to the changing field. Formally, one can express this condition as \cite{bruno}
\begin{equation}
\hbar R_q \ll\Delta,
\label{ad}
\end{equation}
where $\Delta$ is the spin splitting between the two eigenstates of the interaction Hamiltonian. For $\vec{k}$-cubic Dresselhaus coupling, $\Delta\sim\eta k^3$. Since the electron momenta along the $\hat{x}$ and $\hat{y}$ directions are invariant with respect to time (in the ballistic limit), the variation rate of the effective magnetic field depends only on $k_z$. We estimate $R_q$ from $|\dot{\vec{B}}_D| = |\partial{\vec{B}_D}/\partial{k_z}||\partial{k_z}/\partial{t}| =  |\partial{\vec{B}_D}/\partial{k_z}| (e E_z/\hbar) \approx \delta|\vec{B}_D| /\delta t \approx  \delta|\vec{B}_D|R_q$ where $\delta|\vec{B}_D|\sim k^3$ is the change of magnetic field magnitude, and $|\partial{\vec{B}_D}/\partial{k_z}|\sim k^2$. For an initial momentum $k=k_0$, one then obtains $e E_z \ll \eta k_0^4$. Assuming $k_0 \sim k_F$, we have the adiabaticity condition of
\begin{equation}
E_z\ll \frac{\eta k_F^4}{e}.
\label{adreq}
\end{equation}
This condition can be understood from a simple qualitative picture: for sufficiently small carrier accelerations, the internal magnetic field varies smoothly enough such that the spins adiabatically follow its direction \cite{engel}. In realistic zinc-blende semiconductors such as III-V compound semiconductors, the Dresselhaus coupling parameter ranges from $\eta\approx 25 \text{ eV\AA}^3$ for GaAs to $\eta\approx 220\text{ eV\AA}^3$ for InSb \cite{perel,bernevig2004}. The Fermi wavevector in such systems is $k_F=10^8 \text{ m}^{-1}$ for typical doping densities of around $n=10^{16}\text{ cm}^{-3}$ \cite{kato}. Inputting the material values for GaAs into Eq.\ \eqref{adreq} gives the requirement for the applied electric field $E_z \ll E_{\text{ad.}}\sim 10^2 \text{ V/m}$ for the adiabatic limit. In comparison, Kato \emph{et al.}\ \cite{kato} experimentally studied the spin-Hall effect in $n$-doped bulk GaAs and InGaAs samples under an applied field of $E = 10^4$ V/m. Based on our estimate above, the $\vec{E}$ field used by the authors in Ref.\ \cite{kato} appears too large to satisfy the adiabatic condition \eqref{adreq}. Furthermore Ref.\ \cite{kato} does not impose waveguided transport of carriers. So although the system studied in Ref.\ \cite{kato} does not warrant a direct comparison with our theoretical predictions above, we propose that for slight modifications to the experimental setup such that the adiabatic regime is achieved, a detailed analysis of the resulting SHE should consider possible contributions arising from our mechanism.
\subsection{Discussions}
The equations of motion in Eq.\ \eqref{xyz} are valid within time $t<t_s$, where $t_s$ is of the order of typical scattering times governing ballistic transport. In the context of the SHE the discussion of impurities is important as the braking effect of scattering, which is required for the system to reach a steady state, reduces the intrinsic spin-Hall conductivity and in some cases completely destroys it. For example, it is well known that the vertex correction in the Rashba system \cite{sinova} exactly cancels the predicted universal spin-Hall conductivity of $e/8\pi$ \cite{inoue}. This cancellation, however, is a special case for the $\vec{k}$-linear Rashba and Dresselhaus Hamiltonians in 2DEGs and is not the case for general spin-orbit Hamiltonians, e.g.\ in $k^2$-coupling in two dimensional hole gases \cite{bab}, $p$-type bulk semiconductors \cite{mura} and $n$-type bulk semiconductors with $k^3$-Dresselhaus SOC \cite{bernevig2004} (the present system). Our intrinsic spin-Hall conductivity is therefore expected to survive even in the presence of impurities.\\
We compute the predicted value for the spin-Hall conductivity $\sigma_s$ in GaAs to be $\sigma_s \approx 19.4\text{ }\Omega^{-1}\text{m}^{-1}$ (assuming $\lambda=3$ and normalized to have units of charge conductivity) for a ballistic sample. From Ref.\ \cite{kato}, and taking into consideration the constrained Fermi sphere due to the waveguide, the longitudinal charge conductivity is estimated to be $\sigma_c \approx 167\text{ }\Omega^{-1}\text{m}^{-1}$, so our spin-Hall effect can be fairly prominent in the background of the charge conduction. As shown in Fig.\ \ref{spin_cond.fig}, even though $\sigma_s$ drops off rapidly with increasing $\lambda$, the rate of decrease of the ratio $\sigma_s/\sigma_c$ is weaker, so this remains true for larger $\lambda$.
 One should note however that because of the definition of spin current used, our value for $\sigma_s$ cannot be directly related to the observed spin accumulation in actual samples, in which the spin is not a good quantum number because of SOC. To do so, one should use the alternative, conserved spin current definition \cite{shi}. Nevertheless, the simpler conventional definition provides a useful insight and working order of magnitude for $\sigma_s$.\\
The mechanism for the SHE described in this article should be contrasted from the intrinsic mechanisms described in Ref.\ \cite{bernevig2004} for $n$-type bulk semiconductors, and in Ref.\ \cite{sinova} for $n$-doped 2 dimensional electron systems with Rashba SOC, as we explain below. The latter mechanisms for SHE can be viewed as a spin precessional (``torque-based'') effect about the spin-orbit field in the presence of an applied $\vec{E}$ field. As we have seen, the drift action of $\vec{E}$ affects the spin-orbit field $\vec{B}(\vec{k})$. As explained in our paper, for small electric fields the spins adiabatically follow the direction of $\vec{B}(\vec{k})$, but there is also an accompanying non-adiabatic correction to the effective field experienced by electrons \cite{engel, aharonov}. This component has the form $(\dot{\vec{B}}\times\vec{B})/|B|^2$ and arises from the time-dependence of $\vec{B}(\vec{k})$. The SHEs of Refs.\ \cite{bernevig2004,sinova} occur as a result of the spin-precession about this field. Because of SOC, the precession behavior (rotation) is governed by the electron momentum i.e.\ the spins of electrons traveling in opposite transverse directions tilt in an antiparallel manner, resulting in a finite spin-Hall conductivity. On the other hand, in the topological effect induced by the adiabatic relaxation of spins to $\vec{B}(\vec{k})$ described presently, electrons in antiparallel spin states experience opposite anomalous transverse velocities, also giving rise to non-zero $\sigma_s$ (the effect is ``force-based''). The physical origin of the Lorentz-type force which gives rise to the anomalous velocities is, however, related to the non-adiabatic correction of the spin-orbit field \cite{aharonov}.\\
We briefly discuss a possible experimental setup for the detection of our effect, that is similar to that used in Ref.\ \cite{kato}, but (i) with an applied electric field which guarantees adiabaticity as in Eq.\ \eqref{adreq}, and (ii) an implementation of a waveguide structure for the carrier transport e.g.\ through the use of electrostatic gates. Furthermore, the Hall bar should ideally have longitudinal dimension that is of the order of the mean free path ($\Lambda$), to impose ballistic transport of the carriers; for larger samples the effect of impurities will reduce the spin-Hall conductivity. At low temperature, $n$-type bulk GaAs has $\Lambda = 0.1-10 \mu$m depending on the quality of the sample and the doping concentration, (the longitudinal length of the Hall bar in Ref.\ \cite{kato} was $300\mu$m). Without any Hall contacts attached to the sample, the constant spin supply from our effect should be manifested as spin accumulation at the sample edges. Since the spin relaxation time of conduction electrons in semiconductor systems is quite long ($\sim 100$ ps), the spatial distribution of the resulting spin accumulation should be detectable by Kerr rotation microscopy techniques. Identification of our effect from other mechanisms including the extrinsic and precessional-intrinsic effects could be made readily based on the knowledge that the contribution to the total spin-Hall conductivity from our effect is sensitive to the applied electric field, the cut-off being at the adiabatic field limit of $E=E_{\text{ad.}}$.\\
In summary we described an intrinsic, topological spin-Hall effect in waveguided, $n$-type bulk zinc-blende semiconductors in the adiabatic applied field limit. The effect may be detected using standard techniques carried out in previous works, and readily distinguished from other known mechanisms of the SHE.

    %
    %

    \end{document}